\documentstyle [12pt,eqsecnum,aps,amsfonts] {revtex}
\input epsf
\newcommand{\beq}{\begin{equation}}
\newcommand{\eeq}{\end{equation}}
\newcommand{\beqs}{\begin{eqnarray}}
\newcommand{\eeqs}{\end{eqnarray}}

\begin{document}
\draft

\baselineskip 6.0mm

\title{Exact Partition Functions for Potts Antiferromagnets on Cyclic 
Lattice Strips}

\author{Robert Shrock$^{(a)}$\thanks{email: robert.shrock@sunysb.edu}
\and Shan-Ho Tsai$^{(b)}$\thanks{email: tsai@hal.physast.uga.edu}}

\address{(a) \ Institute for Theoretical Physics \\
State University of New York \\
Stony Brook, N. Y. 11794-3840}

\address{(b) \ Department of Physics and Astronomy \\
University of Georgia \\
Athens, GA 30602}

\maketitle

\vspace{10mm}

\begin{abstract}

We present exact calculations of the zero-temperature partition function of the
$q$-state Potts antiferromagnet on arbitrarily long strips of the square,
triangular, and kagom\'e lattices with width $L_y=2$ or 3 vertices and with
periodic longitudinal boundary conditions.  From these, in the limit of
infinite length, we obtain the exact ground-state entropy $S_0=k_B \ln W$.
These results are of interest since this model exhibits nonzero ground state
entropy $S_0 > 0$ for sufficiently large $q$ and hence is an exception to the
third law of thermodynamics.  We also include results for homeomorphic
expansions of the square lattice strip.  The analytic properties of $W(q)$ are
determined and related to zeros of the chromatic polynomial for long finite
strips.

\end{abstract}


\vspace{16mm}

\pagestyle{empty}
\newpage

\pagestyle{plain}
\pagenumbering{arabic}
\renewcommand{\thefootnote}{\arabic{footnote}}
\setcounter{footnote}{0}

\section{Introduction}

The $q$-state Potts antiferromagnet (AF) \cite{potts,wurev} exhibits nonzero
ground state entropy, $S_0 > 0$ (without frustration) for sufficiently large
$q$ on a given graph or lattice.  This is equivalent to a ground state
degeneracy per site $W > 1$, since $S_0 = k_B \ln W$.  Such nonzero ground
state entropy is important as an exception to the third law of thermodynamics
\cite{al,cw}.  The zero-temperature partition function of the above-mentioned
$q$-state Potts antiferromagnet on a graph $G$ satisfies
\beq
Z(G,q,T=0)_{PAF}=P(G,q)
\label{zp}
\eeq
where $P(G,q)$ is the chromatic polynomial
expressing the number of ways of coloring the vertices of the graph $G$ with
$q$ colors such that no two adjacent vertices have the same color
\cite{rrev,rtrev}. Thus, formally, 
\beq
W(\{G\},q) = \lim_{n \to \infty} P(G,q)^{1/n}
\label{w}
\eeq
where $n=v(G)$ is the number of vertices of $G$, and we use the notation
\beq
\{G\} = \lim_{n \to \infty}G
\label{ginf}
\eeq 
The minimum number of colors needed for this
coloring of $G$ is called its chromatic number, $\chi(G)$.
At certain special points $q_s$ (typically $q_s=0,1,.., \chi(G)$),
one has the noncommutativity of limits
\beq
\lim_{q \to q_s} \lim_{n \to \infty} P(G,q)^{1/n} \ne \lim_{n \to \infty}
\lim_{q \to q_s}P(G,q)^{1/n}
\label{wnoncom}
\eeq
and hence it is necessary to specify the order of the limits in the
definition of $W(\{G\},q_s)$ \cite{w}.  As in \cite{w}, we shall use the first
order of limits here; this has the advantage of removing certain isolated
discontinuities that could be present in the $W$ with the opposite order of
limits.  In addition to Refs. \cite{wurev}-\cite{w}, some other previous
related works include Refs. \cite{lieb}-\cite{pg}.  Since $P(G,q)$ is a
polynomial, one can generalize $q$ from ${\mathbb Z}_+$ to ${\mathbb R}$ and
indeed to ${\mathbb C}$.  $W(\{G\},q)$ is a real analytic function for real $q$
down to a minimum value, $q_c(\{G\})$ \cite{w,p3afhc}.  For a given $\{G\}$, we
denote the continuous locus of non-analyticities of $W$ as ${\cal B}$.  This
locus ${\cal B}$ forms as the accumulation set of the zeros of $P(G,q)$ (called
the chromatic zeros of $G$) as $n \to \infty$ and satisfies ${\cal B}(q)={\cal
B}(q^*)$.  A fundamental question concerning the Potts antiferromagnet is the
form of this locus for a given graph family or lattice and, in particular, the
maximal region in the complex $q$ plane to which one can analytically continue
the function $W(\{G\},q)$ from physical values where there is nonzero ground
state entropy, i.e., $W > 1$.  We denote this region as $R_1$. Further, we
denote as $q_c$ the maximal point where ${\cal B}$ intersects the real axis,
which can occur via ${\cal B}$ crossing this axis or via a line segment of
${\cal B}$ lying along this axis.

Although $W(\{G\},q)$ has been calculated exactly for the triangular lattice
\cite{baxter} and a number of different families of graphs \cite{w},
\cite{wc}-\cite{wa2}, it has never been calculated exactly (as a general
function of $q$) for any other 2D (or higher-dimensional) lattice.
Accordingly, one useful procedure is to calculate $W$ exactly on
infinite-length, finite-width strips of various 2D lattices.  As the width
increases, the values of the resulting $W$ functions approach those of the $W$
functions on the respective 2D lattices.  We have carried out this procedure
for strips with free boundary conditions in the longitudinal direction (in
which the length goes to infinity) and free and periodic boundary conditions in
the transverse direction \cite{strip,w2d}.  For a given lattice, the
calculation of $P$ and $W$ on a cyclic strip of a lattice, i.e a strip with
periodic boundary conditions in the longitudinal direction (and some choice of
boundary conditions in the transverse direction) is more difficult and the
results more complicated than is the case for the corresponding strip with free
longitudinal boundary conditions (and the same transverse boundary conditions).
Despite the greater difficulty, the calculations of $W$ on cyclic strip graphs
are worth the effort, because, as explained further below, the resultant $W$
functions exhibit, for finite $L_y$, certain analytic features expected for the
$W$ functions on the infinite 2D lattice, whereas for strips with free
longitudinal boundary conditions, these features are absent (approached in the
limit $L_y \to \infty$).

Accordingly, we have carried out exact calculations of the chromatic
polynomials $P$ and, in the infinite-length limit, the resultant functions $W$,
for cyclic strips of the square and kagom\'e lattices, and we report the
results here.\footnote{\footnotesize{A brief report of some of the results for
$W$ is given in Ref. \cite{wcyl}.}}  Since the $W$ functions that we have
computed for these strips exhibit the analytic properties expected for the $W$
functions on the respective 2D lattices, they constitute, in this sense, the
closest exact results that one has to these $W$ functions.  We also include
results for some other families of cyclic strips.  For all of these strips the
boundary conditions in the transverse direction are free.

 From our earlier exact calculations of $W$ on a number of
families of graphs we have inferred several general results on ${\cal B}$: (i)
for a graph $G$ with well-defined lattice structure, a sufficient condition
for ${\cal B}$ to separate the $q$ plane into different regions is that $G$
contains at least one global circuit, defined as a route following a lattice
direction which has the topology of $S^1$ and a length $\ell_{g.c.}$ that goes
to infinity as $n \to \infty$ \cite{strip}.\footnote{\footnotesize{Some 
families of graphs that do not have regular lattice directions have 
noncompact loci ${\cal B}$ that separate the $q$ plane into different 
regions \cite{wa,wa3,wa2}.}}

  For a $d$-dimensional lattice graph, the existence of global circuits is
equivalent to having periodic boundary conditions (BC's) in at least one
direction. Further, (ii) the general condition for a family $\{G\}$ to have a
locus ${\cal B}$ that is noncompact (unbounded) in the $q$ plane \cite{wa}
shows that a sufficient (not necessary) condition for $\{G\}$ to have a
compact, bounded locus ${\cal B}$ is that it is a regular lattice
\cite{wa,wa3,wa2}.  The third and fourth general features are that for graphs
that (a) contain global circuits, (b) cannot be written in the form $G=K_p + H$
\cite{wc}\footnote{\footnotesize{The complete graph on $p$ vertices, denoted
$K_p$, is the graph in which every vertex is adjacent to every other vertex.
The ``join'' of graphs $G_1$ and $G_2$, denoted $G_1 + G_2$, is defined by
adding bonds linking each vertex of $G_1$ to each vertex in $G_2$.  Graph
families with ${\cal B}$ not including $q=0$ are given in
\cite{wc,wa,wa3,wa2}.}}, and (c) have compact ${\cal B}$, we have observed that
${\cal B}$ (iii) passes through $q=0$ and (iv) crosses the positive real axis,
thereby always defining a $q_c$.  As noted, the advantage of cyclic strip
graphs is that these properties are present for each finite $L_y$ rather than
only being approached in the limit $L_y \to \infty$ as for open strips.

Our present results also strengthen the evidence for our earlier conjectures
\cite{strip,wa2} that on a graph with well-defined lattice structure, a
necessary property for there to be chromatic zeros and, in the $n \to \infty$
limit, a locus ${\cal B}$ including support for $Re(q) < 0$, is that the graph
has at least one global circuit. (This is known not to be a sufficient
property, as shown, e.g., by the circuit and ladder graphs, which have such
global circuits, but whose chromatic zeros and loci ${\cal B}$ have support
only for $Re(q) \ge 0$.)  

\section{General Properties and Calculational Method} 

   A general form for the chromatic polynomial of an $n$-vertex strip graph 
$G$ of the type considered here (and more generally, a graph built up
recursively by the successive addition of a given subgraph) is 
\beq
P(G,q) =  \sum_{j=1}^{N_a} c_j(q)(a_j(q))^{t_j n}
\label{pgsum}
\eeq 
where $c_j(q)$ and $a_j(q)$ are certain functions of $q$. Here the
$a_j(q)$ and $c_j(q)$ are independent of $n$, and $t_j$ is a $G$-dependent
constant.\footnote{\footnotesize{The labelling convention for the summation
index $j$ in eq. (\ref{pgsum}) is the same as that in our Ref. \cite{hs}; this
differs from the convention we used in Ref. \cite{w}, where $j$ started at 0,
and where the term denoted here as $c_1(a_1)^{t_1n}$ was labelled simply as
$c_0$ since $c_0$ was either 1 or proportional to $(-1)^n$, and hence did not
grow with $n$ like $(const.)^n$ with $|const.| > 1$.}}  As before \cite{w}, we
define a term $a_\ell(q)$ as ``leading'' ($\ell$) if it dominates the $n \to
\infty$ limit of $P(G,q)$; in particular, if $N_a \ge 2$, then it satisfies
$|a_\ell(q)| > |a_j(q)|$ for $j \ne \ell$, so that
$|W|=|a_\ell|^{t_\ell}$.  The locus ${\cal B}$ occurs where there is a
nonanalytic change in $W$ as the leading terms $a_\ell$ in eq. (\ref{pgsum})
changes.  Note that a term of the form $a_1=\pm 1$ may be
absent.  For some families of graphs, the $c_j(q)$ and $a_j(q)$ are
polynomials.  However, there are also many families of graphs for which
$c_j(q)$ and $a_j(q)$ are not polynomial, but instead, are algebraic, functions
of $q$; for these families the property that the chromatic polynomial is, in
fact, a polynomial of $q$, is not manifest from the expression (\ref{pgsum}).

For a strip graph $G_s$ our calculational method is to use the
deletion-contraction theorem \cite{rrev} iteratively to obtain a set
of linear equations which we then solve, to get a generating function
$\Gamma(G_s,q,x)$; the chromatic polynomials for the strip of length $L_x=m$
are then determined as the coefficients in a Taylor series expansion of this
generating function in an auxiliary variable $x$ about $x=0$:
\beq
\Gamma(G_s,q,x) = \sum_{m=m_0}^{\infty}P((G_s)_m,q)x^{m-m_0}
\label{gamma}
\eeq
where $m_0$ depends on the type of strip graph $G_s$ and is naturally chosen as
the minimal value of $m$ for which the graph is well defined.  
The generating functions $\Gamma(G_s,q,x)$ are rational functions of the form
\beq
\Gamma(G_s,q,x) = \frac{{\cal N}(G_s,q,x)}{{\cal D}(G_s,q,x)}
\label{gammagen}
\eeq
with
\beq
{\cal N}(G_s,q,x) = \sum_{j=0}^{d_{\cal N}} A_{G_s,j}(q) x^j
\label{n}
\eeq
and
\beq
{\cal D}(G_s,q,x) = 1 + \sum_{j=1}^{d_{\cal D}} b_{G_s,j}(q) x^j
\label{d}
\eeq
where the $A_{G_s,i}$ and $b_{G_s,i}$ are polynomials in $q$ (with no common
factors) \footnote{\footnotesize{As in Ref. \cite{hs}, we use the notation 
$A_{G_s,i}$ rather than our earlier notation $a_{G_s,i}$ in Ref. \cite{strip} 
in order to avoid confusion with the terms $a_j$ in the general formula 
(\ref{pgsum}).}}, 
and the degrees of the numerator and denominator, as polynomials in the 
auxiliary variable $x$, are
\beq
d_{\cal N} = deg_x({\cal N})
\label{dn}
\eeq
and 
\beq
d_{\cal D} = deg_x({\cal D})
\label{dd}
\eeq
Writing the denominator
of the generating function $\Gamma(G_s,q,x)$ in factorized form, we have
\beq
{\cal D}(G_s,q,x) = \prod_{j=1}^{d_{\cal D}}(1-\lambda_{G_s,j}(q)x)
\label{lambdaform}
\eeq 
(c.f. eq. (4.2) in Ref. \cite{strip} or (2.9) in Ref. \cite{hs}). 
The chromatic polynomials can equivalently be calculated from a corresponding
recurrence relation (see eq. (2.25) in Ref. \cite{hs}). 

We define the polynomial
\beq
D_k(q) = \frac{P(C_k,q)}{q(q-1)} = a^{k-2}\sum_{j=0}^{k-2}(-a)^{-j} =
\sum_{s=0}^{k-2}(-1)^s {{k-1}\choose {s}} q^{k-2-s}
\label{dk}
\eeq
where $a=q-1$ and $P(C_k,q)$ is the well-known chromatic polynomial for 
the circuit (cyclic) graph $C_k$ with $k$ vertices,
\beq
P(C_k,q) = a^k + (-1)^ka
\label{pck}
\eeq

In Table 1 we list the values of $d_{\cal N}$ and $d_{\cal D}$ for various
families of lattice strip graphs considered in this paper and in our previous
ones.  The form of ${\cal N}$ and value of $d_{\cal N}$, (but not ${\cal D}$ or
$d_{\cal D}$) depend on one's convention for $m_0$ in eq. (\ref{gamma}); the
values given in Table 1 correspond to the conventions that we use in this paper
and our previous ones. 

In eq. (2.17) of Ref. \cite{hs} we gave a general formula that enables one
to calculate the chromatic polynomial in the closed form (\ref{pgsum}) from the
generating function or vice versa.  Since the denominators of generating
functions for wider strips of various lattices $G_s$ yield $\lambda_{G_s,j}$'s
in eq. (\ref{lambdaform}) that are algebraic, but not polynomial, functions of
$q$, it can be more convenient to obtain the chromatic polynomials directly 
 from
the expansion of the generating functions.  However, for the simplest types of
strips, the $\lambda_{G_s,j}$'s are polynomial, and in these cases the 
closed-form
expression (\ref{pgsum}) is the most convenient.  In the case of cyclic strips
of the square lattice, it is worthwhile to give the generating functions for
the simpler cases $L_y=1$ (the circuit graph) and $L_y=2$ (the cyclic ladder
graph) as a comparison with the much more complicated results for the $L_y=3$
case that we are reporting here.  For these cyclic strips one can write 
\beq
\Gamma(sq(L_y,cyc.),q,x) = \sum_{m=2}^{\infty}P(sq(L_y,cyc.)_m,q)x^{m-2}
\label{gammasqly}
\eeq
As indicated, it is natural to take $m_0=2$ for the cyclic strip of the square
lattice (and for the cyclic strip of the triangular lattice to be discussed
later) because the lowest value of $m$ is $m=2$, for which the respective
strips reduce to a single column of squares or triangles.  
For the minimal case, $L_y=1$, the generating function that yields, as the 
$x^{m-2}$ coefficient, the chromatic polynomial $P(C_m,q)$, is 
\beq
\Gamma(sq(L_y=1,cyc.),q,x) = \frac{q(q-1)}{[1-(q-1)x](1+x)}
\label{gammasqly1}
\eeq
As an example of the $m_0$-dependence of ${\cal N}$, we note that if one used a
definition with $m_0=0$, i.e., 
$\tilde \Gamma(sq(L_y,cyc.),q,x) = \sum_{m=0}^{\infty}P(sq(L_y,cyc.)_m,q)x^m$, 
(where the $m=0,1$ coefficients are only formal) then 
\beqs
\tilde \Gamma(sq(L_y=1,cyc.),q,x) & = & \frac{q[1-(q-2)x]}{[1-(q-1)x](1+x)} 
= \frac{1}{1-(q-1)x} + \frac{q-1}{1-x} \cr\cr
& = & \sum_{m=0}^\infty \Bigl [ (q-1)^m + (q-1)(-1)^m \Bigr ] x^m 
\label{gammasqly1m00}
\eeqs
which, of course, agrees with the expression for $P(sq(L_y=1)_m,q) \equiv 
P(C_m,q)$ given in eq. (\ref{pck}). 
For $L_y=2$, from the known chromatic polynomial \cite{bds}, we determine the
corresponding generating function for eq. (\ref{gammasqly}) (with the
convention $m_0=2$) to be
\beq
\Gamma(sq(L_y=2,cyc.),q,x) = \frac{q(q-1)[D_4 -2(q-1)x-(q-1)(q-3)D_4x^2]}
{(1-D_4x)(1-x)[1-(1-q)x][1-(3-q)x]}
\label{gammasqly2}
\eeq
where $D_4=q^2-3q+3$.  As will be seen from our new results, the complexity of
the generating function increases considerably when one goes from width $L_y=2$
to $L_y=3$.

\begin{table}
\caption{
Properties of generating functions for strip graphs $G_s$. }
\begin{center}
\begin{tabular}{ccc}
$G_s$ & $deg_x({\cal N})$ & $deg_x({\cal D})$ \\
\hline 
$sq(L_y=1,FBC_x)$   & 0 & 1 \\
$sq(L_y=2,FBC_x)$   & 0 & 1 \\
$sq(L_y=3,FBC_x)$   & 1 & 2 \\
$sq(L_y=4,FBC_x)$   & 2 & 3 \\ 
\hline
$sq(L_y=1,PBC_x)$   & 0 & 2 \\
$sq(L_y=2,PBC_x)$   & 2 & 4 \\
$sq(L_y=3,PBC_x)$   & 8 & 10 \\
\hline
$t(L_y=2,FBC_x)$    & 0 & 1 \\
$t(L_y=3,FBC_x)$    & 1 & 2 \\
$t(L_y=4,FBC_x)$    & 3 & 4 \\ 
\hline
$t(L_y=2,PBC_x)$    & 2 & 4 \\
\hline
$kg(L_y=2,FBC_x)$   & 1 & 2 \\
\hline
$kg(L_y=2,PBC_x)$   & 8 & 9 \\
\hline
\end{tabular}
\end{center}
\label{degtable}
\end{table}

For a given $G_s$, in region $R_1$, the function $W$ is given by 
\beq
W=(\lambda_{max})^t
\label{wr1}
\eeq
For $q$ in regions other than $R_1$, one can only determine the magnitude 
$|W(\{G\},q)|$ unambiguously \cite{w}; this is given by 
\beq
|W|=|\lambda_{max}|^t
\label{wmag}
\eeq
where $\lambda_{max}$ denotes the $\lambda$ with maximal magnitude in the
respective region and 
\beq
t = \frac{m}{n}
\label{t}
\eeq 
For the square and triangular strips, $t=1/L_y$ while for the kagom\'e
strip considered here, $t=1/5$. 
Thus, $W$ and ${\cal B}$ are determined by ${\cal D}$, independent of 
${\cal N}$.  Of course, the determination of ${\cal N}$ is necessary to
calculate the actual chromatic polynomials for particular finite-length
strips.  Besides their intrinsic interest in the context of mathematical graph
theory, these chromatic polynomials for finite strips are important because
from them one can calculate their zeros and thereby check how rapidly these 
approach the asymptotic $n \to \infty$ accumulation set ${\cal B}$.  (Indeed,
for some other families of graphs, we have shown that the chromatic zeros lie 
precisely on the asymptotic loci even for finite $n$ \cite{wc}.)

The organization of the paper is as follows.  Our calculations for the strips
of the square, kagom\'e, and triangular lattices are given in sections 3-5 (see
also the appendix).  In section 6 we include results for certain homeomorphic
expansions of square strips.  Our concluding remarks are given in section 7.

\section{Cyclic Strip of the Square Lattice with $L_{\lowercase{y}}=3$}

We consider the cyclic strip of the square (sq) lattice of length $L_x=m$
vertices and width $L_y=3$ vertices.  This has $n=L_xL_y=3m$ vertices, 
$e=5m$ edges (for the nondegenerate cases $m > 2$), and chromatic number 
(for any $L_y$) 
\beq
\chi(sq(L_y)_m,cyc.) = \cases{ 2 & for even  $m$ \cr
               3 & for odd $m$ \cr }
\label{chisqly3}
\eeq
(For the minimal case $m=2$, the periodic longitudinal boundary conditions lead
the strip to degenerate to a single tower of two squares.) 
We use a generating function of the form 
\beq
\Gamma(sq(L_y=3,cyc.),q,x) = \sum_{m=2}^{\infty} P(sq(L_y=3,cyc.)_m,q)x^{m-2} 
\label{gammasqly3}
\eeq
where $P(sq(L_y=3,cyc.)_m,q)$ is the chromatic polynomial for the strip of
length $m$ vertices.  We obtain a generating function of the form 
(\ref{gammagen}) with (\ref{n}) and (\ref{d}), where 
$deg_x({\cal N})=8$ and $deg_x({\cal D}) = 10$.  Clearly this is 
considerably more complicated than the generating function for the strip with
the same width, $L_y=3$, but with free longitudinal boundary conditions (and
again free transverse boundary conditions), which had 
$deg_x({\cal N}) = 1$ and $deg_x({\cal D}) = 2$.  This and other comparisons
are shown in Table 1. 

\begin{figure}
\centering
\leavevmode
\epsfxsize=3.5in
\begin{center}
\leavevmode
\epsffile{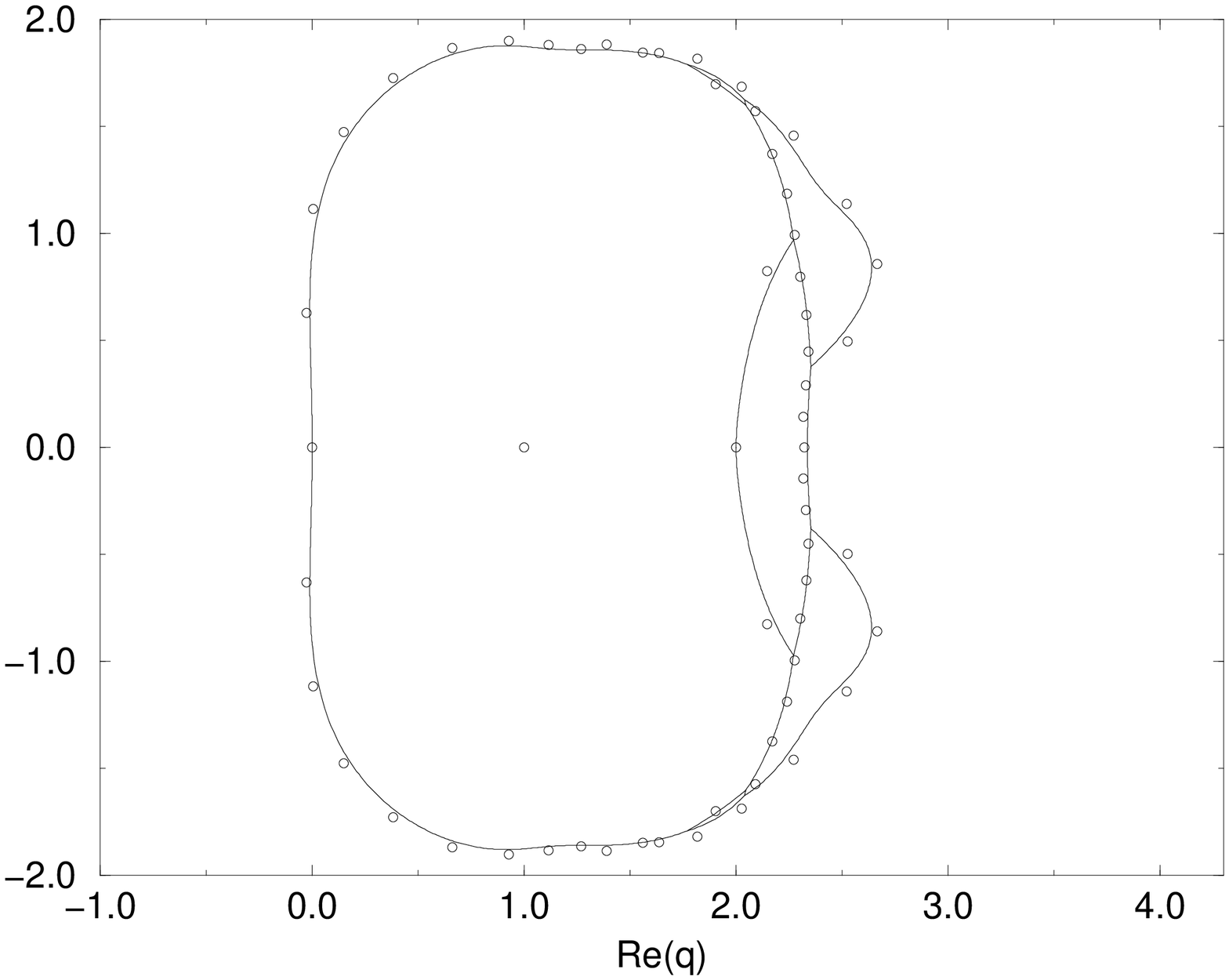}
\end{center}
\vspace{-2cm}
\caption{\footnotesize{Boundary ${\cal B}$ in the $q$ plane for
$W$ function for a cyclic strip of the square lattice with width $L_y=3$.
Chromatic zeros for $m=20$ and thus $n=60$ vertices are shown for comparison.}}
\label{clad3}
\end{figure}

We have calculated the denominator to be 
\beqs
{\cal D}(sq(L_y=3),q,x)&=&[1+x][1-(q-1)x][1-(q-2)x][1-(q-4)x][1+(q-2)^2x]
\cr\cr & & \times [1+b_{sq,11}x+b_{sq,12}x^2]
[1+b_{sq,21}x+b_{sq,22}x^2+b_{sq,23}x^3]
\label{dsqly3}
\eeqs
where
\beq
b_{sq,11}=-(q-2)(q^2-3q+5)
\eeq
\beq
b_{sq,12}=(q-1)(q^3-6q^2+13q-11)
\eeq
\beq
b_{sq,21}=2q^2-9q+12
\eeq
\beq
b_{sq,22}=q^4-10q^3+36q^2-56q+31
\eeq
\beq
b_{sq,23}=-(q-1)(q^4-9q^3+29q^2-40q+22)
\eeq
Thus, 
\beq
{\cal D}(sq(L_y=3,cyc.),q,x)=\prod_{j=1}^{10} (1-\lambda_{sq,j}x)
\label{dsqly3lam}
\eeq
where the first five $\lambda$'s can be read off immediately from the 
linear factors in eq. (\ref{dsqly3}),
\beq
\lambda_{sq,(6,7)} \equiv 
\lambda_{sq,1,\pm} = \frac{1}{2}\biggl [ -b_{sq,11} \pm 
(b_{sq,11}^2-4b_{sq,12})^{1/2} \biggr ]
\label{lambdasq1pm}
\eeq
and $\lambda_{sq,j}$, $j=8,9,10$ are the roots of the cubic
\beq
\xi^3+b_{sq,21}\xi^2+b_{sq,22}\xi+b_{sq,23}=0
\label{sqcubic}
\eeq
 From this we have calculated the
locus ${\cal B}$, shown in Fig. \ref{clad3}.  This locus separates the $q$
plane into seven regions: $R_1$ including the real interval $q > q_c$; 
$R_2$, to
the left of $R_1$; $R_3$ including the real interval $0 < q < 2$; two 
complex-conjugate (c.c.) phases $R_4,R_4^*$ centered at $q \simeq 2.4
\pm 0.9i$; and two additional quite narrow, sliver-like c.c. phases 
$R_5,R_5^*$ centered at $q \simeq 1.95 \pm 1.75i$. In region $R_1$, 
the dominant $\lambda$ is $\lambda_{sq,6} \equiv \lambda_{sq,1,+}$, so that 
\beqs
W(sq(L_y=3,cyc.),q) & = & 2^{-1/3}\biggl [ (q-2)(q^2-3q+5) + \nonumber \\ & &
\Bigl [(q^2-5q+7)(q^4-5q^3+11q^2-12q+8) \Bigr ]^{1/2} \biggr ]^{1/3} \quad {\rm
for} \quad q \in R_1
\label{wsqly3r1}
\eeqs
In region $R_2$, the dominant $\lambda$ is $q-4$, so 
\beq
|W(sq(L_y=3,cyc.),q)|=|q-4|^{1/3} \quad {\rm for} \quad q \in R_2
\label{wsqly3r2}
\eeq
The value of $q_c$ for this strip is given as the real solution to the
equation $|W_{R_1}|=|W_{R_2}|$, which reduces to 
\beq
2q^4-16q^3+51q^2-86q+67=0
\label{qceq}
\eeq
yielding
\beq
q_c(sq(L_y=3,cyc.))=2.33654
\label{qcsqly3}
\eeq 
At this point $W=1.18487$.  In region $R_3$, the dominant $\lambda$ is the
maximal root of the cubic (\ref{sqcubic}).  The degeneracy condition
$|W_{R_2}|=|W_{R_3}|$ yields the value $q=2$ for the point separating $R_3$ and
$R_2$ on the real axis.  Note that the portion of the boundary ${\cal B}$
separating regions $R_2$ and $R_3$ is an arc of the circle $|q-4|=2$. In the
c.c. pairs of regions ($R_4,R_4^*$) and ($R_5,R_5^*$) the dominant $\lambda$'s
are the other two roots of the cubic (\ref{sqcubic}).  As discussed before,
these results depend only on ${\cal D}$.  As is evident in Fig. \ref{clad3},
${\cal B}$ contains several multiple points, forming conjugate complex pairs.
The property that there is no multiple point at $q_c$ follows because only two
of the $\lambda$'s become degenerate in magnitude at this point.  This may be
contrasted with the situation for the $L_y=2$ cyclic strip, where three leading
terms become degenerate, leading to a multiple point, at
$q_c(sq(L_y=2,cyc.))=2$, at which four curves come together \cite{w}.  As noted
above, these results depend only on ${\cal D}$. 

The coefficient functions for the numerator are rather lengthy and hence are
given in the Appendix.  Together with the above results for ${\cal D}$, these
yield, via eq. (\ref{gammasqly3}), the chromatic polynomials for the cyclic
strip of the square lattice of arbitrary length.  From the generating function,
using eq. (2.17) of Ref. \cite{hs} (with $m \to m-1$ to match our current
notational conventions), we calculate the chromatic polynomial
\beqs 
P(& & sq(L_y = 3,cyc.)_m,q) = (q^3-5q^2+6q-1)(-1)^m \cr\cr 
& & + (q^2-3q+1)\Bigl [(q-1)^m+(q-2)^m+(q-4)^m \Bigr ] 
+ (q-1)[-(q-2)^2]^m + \Bigl [ (\lambda_{sq,6})^m+(\lambda_{sq,7})^m \Bigr ]
\cr\cr & & +
(q-1)\Bigl [(\lambda_{sq,8})^m+(\lambda_{sq,9})^m+(\lambda_{sq,10})^m \Bigr ]
\label{psqly3}
\eeqs 
This can also be derived directly from ${\cal D}$ without using ${\cal N}$ by
matching with calculations of $P$ for the first few values of $m$.  Thus, in
the notation of eq. (2.17) of Ref. \cite{hs} or eq. (\ref{pgsum}) here,
$a_{sq,j}=\lambda_{sq,j}$, $c_{sq,1}=q^3-5q^2+6q-1$, $c_{sq,j}=q^2-3q+1$ for
$j=2,3,4$, $c_{sq,j}=q-1$ for $j=5,8,9,10$, and $c_{sq,j}=1$ for $j=6,7$.  In
Fig. \ref{clad3} we show chromatic zeros for a strip of length $m=20$ and thus
$n=60$ vertices, for comparison with the $n \to \infty$ accumulation set ${\cal
B}$.  Besides the fact that these zeros lie close to ${\cal B}$, we observe
that the density is largest on the right-hand side of the ``main'' curve in
${\cal B}$ and is rather small for the curve bounding the region $R_2$ on the
left.

An interesting feature of this family of graphs is that the chromatic zeros and
their accumulation set ${\cal B}$ have support for $Re(q) < 0$.  Chromatic
zeros with negative real parts occur first for $m=10$, i.e., $n=30$. 

\vspace{2mm}

\section{Cyclic Strip of the Kagom\'e Lattice}

We next consider a cyclic strip of the kagom\'e lattice comprised of $m$
hexagons with each pair sharing two triangles as adjacent polygons (as shown
for the open strip in Fig. 1 in Ref. \cite{strip}).   This has $n=5m$ vertices
and $e=8m$ edges.  The chromatic number is $\chi=3$. We use a generating 
function of the form
\beq
\Gamma(kg(L_y=2,cyc.),q,x) = \sum_{m=1}^{\infty} P(kg(cyc.)_m,q)x^{m-1} 
\label{gammakg}
\eeq
where $P(kg(cyc.)_m,q)$ is the chromatic polynomial for the cyclic strip with 
$m$ hexagons.  The generating function $\Gamma$ has 
$deg_x({\cal N})=8$ and $deg_x({\cal D})=9$, and is thus much more complicated
than the generating function for the kagom\'e strip of the same width with 
free longitudinal boundary conditions, which had 
$deg_x({\cal N})=1$ and $deg_x({\cal D})=2$ (see Table 1). 
We have calculated 
\beqs
{\cal D}(kg(L_y,cyc.),q,x) &=& 
(1+b_{kg,11}x+b_{kg,12}x^2)(1+b_{kg,21}x+b_{kg,22}x^2)
(1+b_{kg,31}x+b_{kg,32}x^2) \times \cr\cr
& & [1-(q-2)x][1-(q-4)x][1-(q-1)(q-2)^2x] 
\label{dkag}
\eeqs
where
\beq
b_{kg,11}=-(q-2)(q^4-6q^3+14q^2-16q+10)
\eeq
\beq
b_{kg,12}=(q-1)^3(q-2)^3
\eeq
\beq
b_{kg,21}=-q^3+7q^2-19q+20
\eeq
\beq
b_{kg,22}=(q-1)(q-2)^3
\eeq
\beq
b_{kg,31}=11-9q+2q^2
\eeq
\beq
b_{kg,32}=-(q-1)(q-2)^2
\eeq
We thus find, for $1 \le j \le 6$, 
\beq
\lambda_{kg,j} \equiv \lambda_{kg,k,\pm}=\frac{1}{2}\biggl [-b_{kg,k1} \pm 
(b_{kg,k1}^2-4b_{kg,k2})^{1/2} \ \biggr ] \ , \quad k=1,2,3
\eeq
\beq
\lambda_{kg,7}=q-2
\eeq
\beq
\lambda_{kg,8}=q-4
\eeq
\beq
\lambda_{kg,9}=(q-1)(q-2)^2
\eeq

\begin{figure}
\centering
\leavevmode
\epsfxsize=3.5in
\begin{center}
\leavevmode
\epsffile{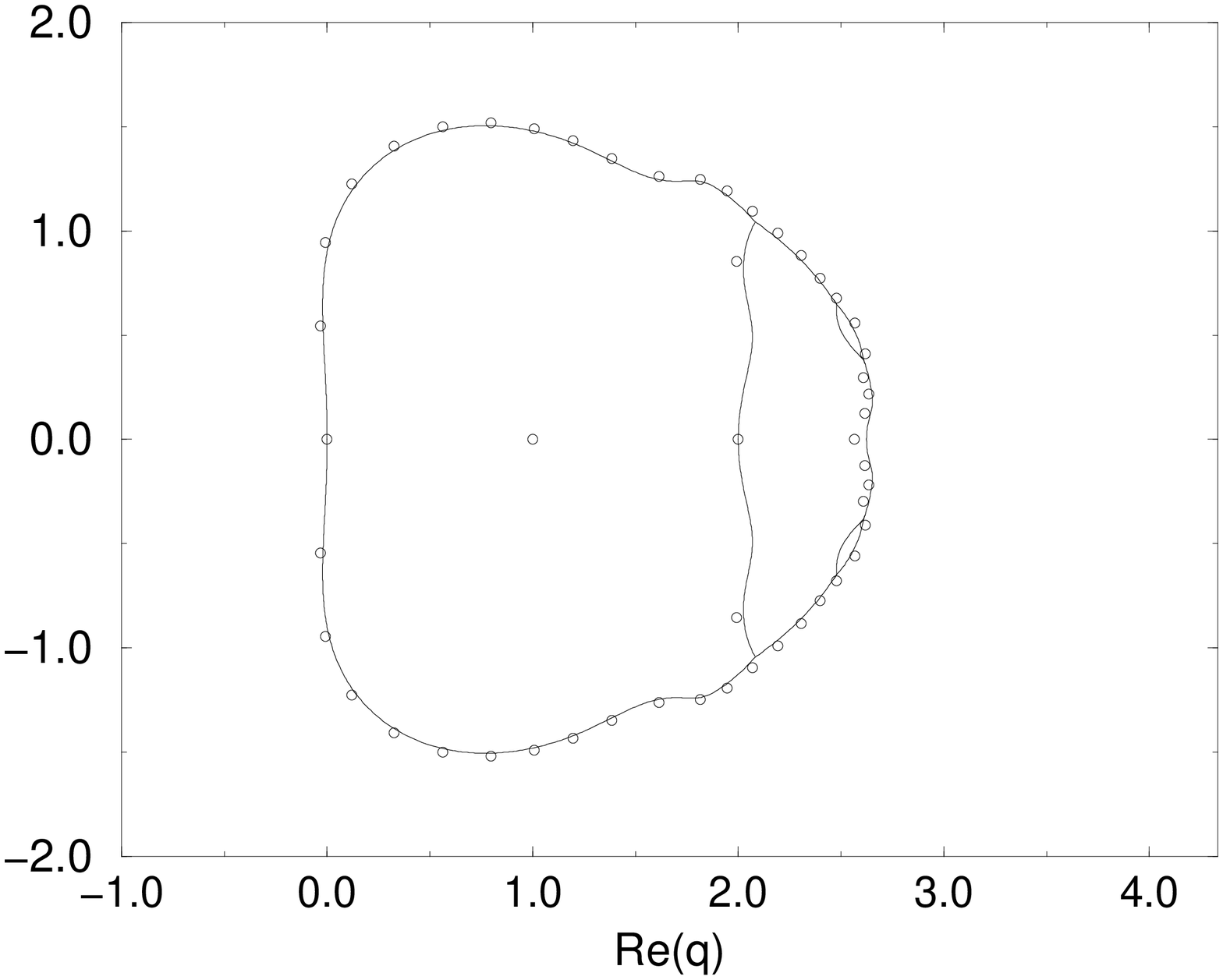}
\end{center}
\vspace{-2cm}
\caption{\footnotesize{Locus ${\cal B}$ for $W$ for $\infty \times 2$ cyclic
strip of the kagom\'e lattice. Chromatic zeros for $m=10$ ($n=50$) are also
shown for comparison.}}
\label{cyckag}
\end{figure}

The coefficient functions for the numerator are rather lengthy and hence are
given in the Appendix.  From the generating function, via eq. (2.17) of
Ref. \cite{hs}, or directly from ${\cal D}$ by matching results with
computations of $P$ for the first few values of $m$, we calculate the chromatic
polynomial
 \beqs 
P(kg(L_y=2)_m,q) &
= & (q^2-3q+1)(q-4)^m+(q-1)\Bigl [(q-1)(q-2)^2 \Bigr ]^m \cr\cr & & + \Bigl
[(\lambda_{kg,1})^m+(\lambda_{kg,2})^m \Bigr ] + (q-1)\Bigl
[(\lambda_{kg,3})^m+(\lambda_{kg,4})^m \Bigr ]
\label{pkag}
\eeqs
There are three $\lambda$'s in ${\cal D}$ that do not appear in $P$, namely,
$q-2$ and $\lambda_{kg,(5,6)} \equiv \lambda_{kg,3,\pm}$.  In the context of
the general formula, eq.  (2.17) of Ref. \cite{hs}, this can be seen as a
result of the vanishing of the corresponding coefficient functions $c_j$ in
eq. (2.19) of that work.  Fig. \ref{cyckag} shows the boundary ${\cal B}$ in
the $q$ plane for the $W$ function.  Again, ${\cal B}$ divides the $q$ plane
into several regions (Fig. 2).  In region $R_1$, $W$ is determined by
$\lambda_{kg,1} \equiv \lambda_{kg,1,+}$:
\beqs
W(\{G_{kg(L_y=2)}\},q) & = & 2^{-1/5}(q-2)^{1/5}\biggl [ q^4-6q^3+14q^2-16q+10
\nonumber \\
& & + \Bigl [q^8-12q^7+64q^6-200q^5+404q^4-548q^3+500q^2-292q+92
  \Bigr ]^{1/2} \biggr ]^{1/5}
\label{wkagw2}
\eeqs
As $q$ decreases through $q_c$, $|W|$ switches to $|q-4|^{1/5}$;
solving the degeneracy equation $|\lambda_{kg,1}|=|4-q|$ yields $q_c=2.62421$
(at which point $W=1.06588$).  This is within about 10 \% of the inferred
exact value $q_c=3$ for the 2D kagom\'e lattice \cite{w2d}.
It is impressive that an infinite strip of width
$L_y=2$ yields a $q_c$ this close to the value for the full 2D lattice.
In the regions including the
intervals $2 < q < q_c$ and $0 < q < 2$, $|W|=|q-4|^{1/5}$ and
$|W|=|\lambda_{kg,4}|^{1/5}$, respectively. It is interesting to
observe that the boundary ${\cal B}$ includes a part with $Re(q) < 0$. Of the
nine coefficient functions in the polynomial in $x$ in ${\cal D}$, the two that
are equal to the coefficient functions for the open strip of the kagom\'e
lattice of the same width are $b_{kg,11}$ and $b_{kg,12}$.  Thus,
the root $\lambda_{kg,1}$ is leading in
region $R_1$ for both open and cyclic longitudinal boundary conditions.

To compare with the asymptotic locus ${\cal B}$, we have calculated the zeros
of the chromatic polynomial for a finite strip with $m=10$, i.e., $n=50$
vertices; these are displayed in Fig.  (\ref{cyckag}).  As in the case of the
$L_y=3$ cyclic strip of the square graph, this cyclic strip of the kagom\'e
lattice has the property that the chromatic zeros and their accumulation set
${\cal B}$ have support for $Re(q) < 0$.  Chromatic zeros with negative real
parts occur first for $m=5$, i.e., $n=25$).

\section{Cyclic Strip of the Triangular Lattice with $L_{\lowercase{y}}=2$}

In this section we consider the cyclic strip of the triangular ($t$) lattice of
length $L_x=m$ vertices and width $L_y=2$ vertices.  This has $n=L_xL_y=2m$
vertices and, for the nondegenerate cases $m \ge 3$, $e=4m$ edges (bonds).  We
consider both the cases of periodic and twisted periodic longitudinal boundary
conditions, and we denote the correponding strips as cyclic and twisted
cyclic (= M\"obius, i.e., opposite ends are identified with reversed 
orientation). For $m=2$ the cyclic and twisted cyclic strips both 
degenerate to the
complete graph on 4 vertices, $K_4$, with $e=6$ edges.  The cyclic strip
has chromatic number 
\beq 
\chi(t(L_y=2,L_x=m),cyc.) = \cases{ 3 & if $m=0$ \
mod 3 \cr 4 & if $m=1$ \ or $m=2$ mod 3 \cr }
\label{chitly2}
\eeq
whereas the twisted cyclic strip has
\beq
\chi(t(L_y=2,L_x=m), \ twisted \ cyc.) = 4
\label{chittly2}
\eeq
For these cyclic strips we use generating functions of the form 
\beq 
\Gamma(t(L_y=2,cyc.),q,x) = \sum_{m=2}^{\infty} P(t(L_y,cyc.)_m,q)x^{m-2} 
\label{gammastart2}
\eeq 
\beq 
\Gamma(t(L_y=2, \ twisted \ cyc.),q,x) = \sum_{m=2}^{\infty} P(t(L_y, \ 
twisted \ cyc.)_m,q)x^{m-2}
\label{gammastart2t}
\eeq
where $P(t(L_y,cyc.)_m,q)$ and 
$P(t(L_y, twisted \ cyc.)_m,q)$ are the chromatic polynomials for the 
respective strips.  For these strips, we find that $deg_x({\cal D})=4$ 
(see Table 1). The degrees in $x$ of the 
numerator for the cyclic and twisted cyclic 
strips are, respectively, $deg_x({\cal N})=2$ and $deg_x({\cal N})=0$.  We 
calculate 
\beq
b_{t,1} = -q^2+6q-10
\label{b1tly1cyc}
\eeq
\beq
b_{t,2} = -(q-3)(2q^2-9q+11)
\label{b2tly1cyc}
\eeq
\beq
b_{t,3} = -(q-2)^2(q^2-6q+10)
\label{b3tly1cyc}
\eeq
\beq
b_{t,4} = (q-2)^4
\label{b4tly1cyc}
\eeq
The denominator of the generating function has the factorized form
\beqs
{\cal D}(t(L_y=2,cyc.),q,x) = & & 
(1-x) \Bigl [ 1-(q-2)^2 x\Bigr ] \Bigl [1+(2q-5)x+(q-2)^2x^2 \Bigr ]
\cr\cr
& & = \prod_{j=1}^4 (1-\lambda_{t,j}x)
\label{dentri}
\eeqs
where 
\beq
\lambda_{t,1} = 1 
\label{lambda1tly2cyc}
\eeq
\beq
\lambda_{t,2} = (q-2)^2
\label{lambda2tly2cyc}
\eeq
\beq
\lambda_{t,3} = \frac{1}{2}\biggl [ 5-2q + \sqrt{9-4q} \ \biggr ]
\label{lambda3tly2cyc}
\eeq
\beq
\lambda_{t,4} = \frac{1}{2}\biggl [ 5-2q - \sqrt{9-4q} \ \biggr ]
\label{lambda4tly2cyc}
\eeq

\begin{figure}
\epsfxsize=3.5in
\begin{center}
\leavevmode
\epsffile{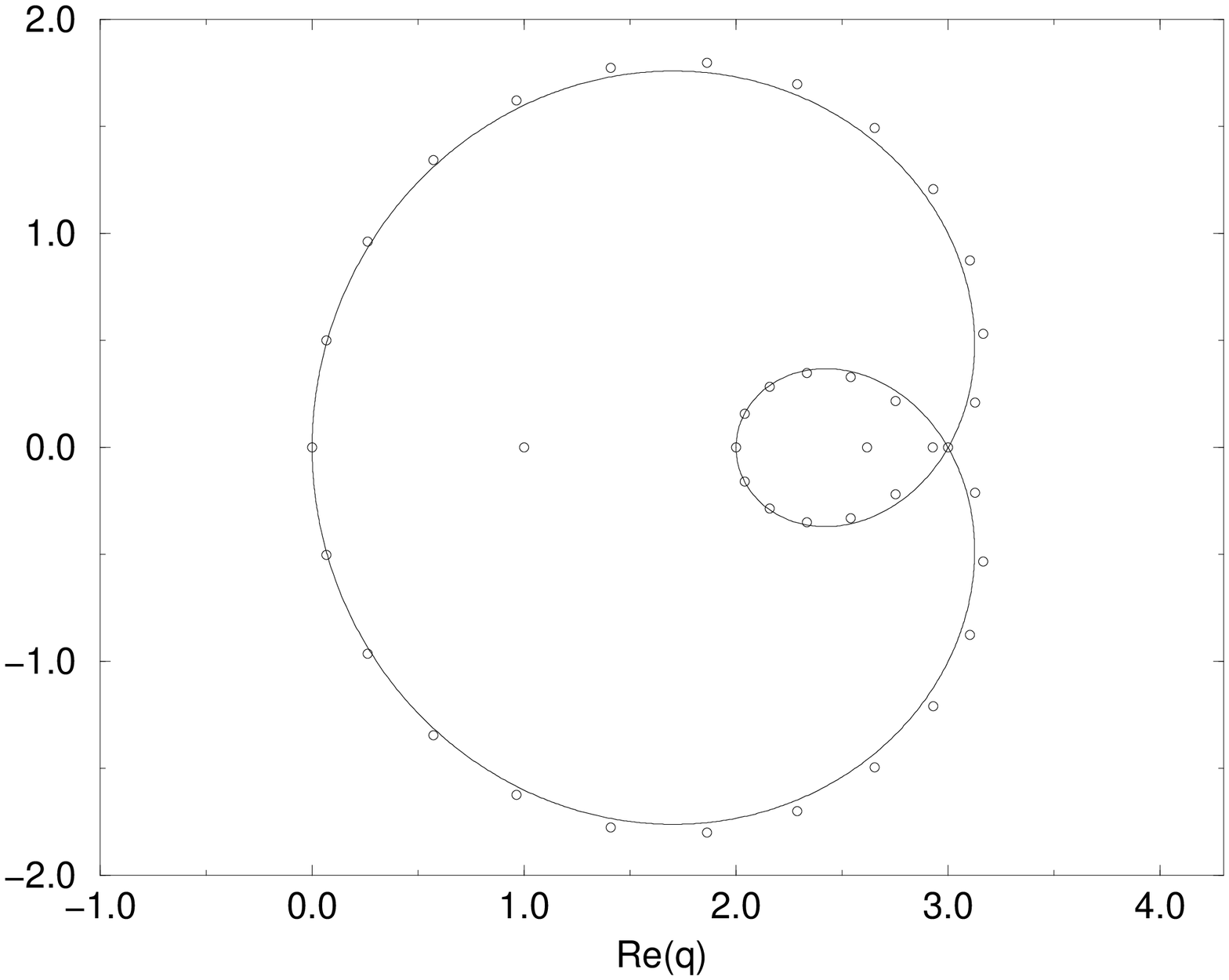}
\end{center}
\vspace{-1 cm}
\caption{\footnotesize{Locus ${\cal B}$ for the function $W$ for a cyclic strip
of the triangular lattice with width $L_y=2$. Chromatic zeros for $m=20$ 
($n=40$ vertices) are shown for comparison.}}
\label{cyctri}
\end{figure}

Because the locus ${\cal B}$ depends only on ${\cal D}$ and is independent of 
${\cal N}$, it is the same for the cyclic and twisted cyclic strips; hence, in
this discussion, we shall group both together and drop the ``twisted'' label. 
The locus ${\cal B}$ is shown in Fig. \ref{cyctri} and separates the complex
$q$ plane into three different regions.  We find that 
\beq
q_c(t(L_y=2,cyc.))=3
\label{qctly2}
\eeq
This may be compared with the value of $q_c$ for the full 2D lattice, which is 
$q_c(tri)=4$ \cite{baxter,w}. 
The region $R_1$ contains the real interval $q > 3$.  The other two regions,
$R_2$ and $R_3$, include the respective real intervals $2 < q < 3$ and 
$0 < q < 2$.  We find that the leading $\lambda's$ in regions
$R_1$, $R_2$ and $R_3$ are respectively $\lambda_{max}=\lambda_2, \ \lambda_1$,
and $\lambda_3$.  It follows that 
\beq
W(t(L_y=2,cyc.),q) = D_3 = q-2 \quad {\rm for} \quad q \in R_1
\label{wtr1}
\eeq
\beq
|W(t(L_y=2,cyc.),q)| = 1 \quad {\rm for} \quad q \in R_2
\label{wtr2}
\eeq
\beq
|W(t(L_y=2,cyc.),q)| = \biggl |\frac{1}{2}\Bigl [ 5-2q + \sqrt{9-4q} \ 
\Bigr ] \biggr |^{1/2} \quad {\rm for}\quad  q \in R_3
\label{wtr3}
\eeq
As an algebraic curve, the boundary ${\cal B}$ has a multiple point of index 
2 at $q=q_c$ in the technical terminology of algebraic geometry \cite{alg}
(i.e. two branches of the curve cross at this point with two separate
tangents).  As is clear in Fig. (\ref{cyctri}), for this strip, ${\cal B}$ has
no support for $Re(q) < 0$.  This is in accord with our earlier inferences
noted above, since the global-circuit condition is necessary, but not
sufficient, for ${\cal B}$ to have support for $Re(q) < 0$. 

For the numerator of the generating function, we find, for the cyclic strip, 
\beq
A_{t,0} = P(K_4,q)=q(q-1)(q-2)(q-3)
\label{a0tly2cyc}
\eeq
\beq
A_{t,1} = q(q-1)(q-2)^2
\label{a1tly2cyc}
\eeq
\beq
A_{t,2} =- q(q-1)(q-2)^4
\label{a2tly2cyc}
\eeq
where $P(K_p,q)$ is the chromatic polynomial for the complete graph on $p$
vertices, 
\beq
P(K_p,q)=\prod_{s=0}^{p-1}(q-s)
\label{pkp}
\eeq
For the twisted cyclic strip (for which $deg_x({\cal N})=0$) we find 
\beq
A_{tt,0}=A_{t,0}=P(K_4,q)
\label{a1ttly2cyc}
\eeq

By the same methods as described above, we obtain 
\beqs
& & P(t(L_y=2,cyc.)_m,q) = q^2-3q+1 + (\lambda_{t,2})^m + (q-1)\biggl [
(\lambda_{t,3})^m + (\lambda_{t,4})^m \biggr ] \cr\cr
& = & q^2-3q+1 + (q-2)^{2m} +
2^{1-m}(q-1)\sum_{s=0}^{[m/2]}{{m}\choose{2s}}(5-2q)^{m-2s}(9-4q)^s
\label{ptly2pol}
\eeqs
\beqs
& & P(t(L_y=2,twisted \ cyc.)_m,q) = -1 + (\lambda_{t,2})^m  
- \frac{(q-1)(q-3)}{\sqrt{9-4q}}\biggl [ 
(\lambda_{t,3})^m - (\lambda_{t,4})^m \biggr ] \cr\cr 
& = & -1 + (q-2)^{2m} - 
2^{1-m}(q-1)(q-3)\sum_{s=0}^{[(m-1)/2]}{{m}\choose{2s+1}}(5-2q)^{m-2s-1}
(9-4q)^s
\label{pttly2pol}
\eeqs
where in the upper limits on the sums, $[x]$ denotes the integral part of $x$.

\section{Asymmetric Homeomorphic Expansions of the Square Strip}

It is also of interest to study homeomorphic expansions of the cyclic square
strip.\footnote{\footnotesize{
A homeomorphic expansion of a graph consists in the addition of
degree-2 vertices to bonds of that graph.}}  In Refs. \cite{hs} we analyzed 
such homeomorphic expansions of a square strip with free longitudinal 
boundary conditions (and free transverse boundary conditions).  In particular,
let us consider the case in which for a $L_y=2$ square strip, we add $k_1-2$ 
vertices to each of the upper horizontal bonds and $k_2-2$ vertices to each 
of the lower horizontal bonds.  For $k_1=k_2$, the resulting $\lambda$'s are
polynomial, and the chromatic polynomials can be written in a relatively simple
form \cite{pg}.  Here we consider the case $k_1 \ne k_2$, which is more
complicated.   We first comment on a wider set of graphs, defined such that for
each square in the original strip one has a choice as to 
whether to assign the $k_1$ vertices to the
upper longitudinal side and $k_2$ vertices to the lower side, or vice versa.
We define a parity vector \beq \Sigma=(\sigma_1,\sigma_2,...\sigma_m)
\label{sigmavector}
\eeq
in which the $\sigma_j=+$ if for the
$j$'th $p$-gon, the assignment is (upper,lower) $=(k_1,k_2)$  and
$\sigma_j=-$ if the assignment is (upper,lower) $=(k_2,k_1)$.
We shall denote this strip graph as
\beq
(Ch)_{k_1,k_2,\Sigma,cyc.,m}=HEL_{k_1-2,k_2-2}\Bigl ( 
G_{sq(L_y=2),cyc.,m}\Bigr ) 
\label{chk1k2smc}
\eeq
It will be convenient to define the special $\Sigma$ vector
\beq
\Sigma_{+,m} = (+,+,...,+)
\label{sigmaplus}
\eeq
where the dimension $m$ of $\Sigma$ can be implicit.
Obviously, the $(Ch)_{k_1,k_2,\Sigma,cyc.,m}$ graphs
are invariant under a reflection about the
longitudinal axis, which in the $k_1 \ne k_2$ case
amounts to a simultaneous reversal of the signs of all
of the $\sigma_j$'s.  Let us define this parity operation on $\Sigma$ as
\beq
P(\Sigma) = -\Sigma
\label{psigma}
\eeq
Then
\beq
(Ch)_{k_1,k_2,\Sigma,cyc.,m}= (Ch)_{k_1,k_2,P(\Sigma),cyc.,m} = 
(Ch)_{k_2,k_1,\Sigma,cyc.,m}
\label{ch1221}
\eeq
The total number of vertices is
\beq
v((Ch)_{k_1,k_2,\Sigma,cyc.,m}) = (p-2)m
\label{vk1k2chain}
\eeq 
independent of $\Sigma$, where
\beq
p=k_1+k_2
\label{pk1k2}
\eeq 
For the strip with free longitudinal boundary conditions, we showed in
Ref.  \cite{hs} that the chromatic polynomial $P((Ch)_{k_1,k_2,\Sigma,m},q)$
was (i) independent of $\Sigma$ and (ii) depended only on the homeomorphic
expansion indices $k_1$ and $k_2$ through their sum, $p$.  However, these
properties do not hold for the cyclic strip $(Ch)_{k_1,k_2,\Sigma,cyc.,m}$.  
This
can be understood as a consequence of the fact that tying together the ends via
the periodic longitudinal boundary conditions increases the constraints on the
coloring of the graphs.

To study asymmetric homeomorphic expansions, we shall consider the case 
$(k_1,k_2)=(k,2)$ (whence $p=k+2$) with $\Sigma=\Sigma_{+,m}$, i.e., the 
family 
\beq
(Ch)_{k_1=k,k_2=2,\Sigma_{+,m},cyc.,m}
\label{k2chain}
\eeq
For this calculation we use a generating function of the form
\beq
\Gamma((Ch)_{k,2,\Sigma_+,cyc.},q,x) = \sum_{m=2}^{\infty} x^{m-2} 
P((Ch)_{k,2,\Sigma_{+,m},cyc.,m},q)
\label{gammask2chain}
\eeq 
 From eq. (\ref{vk1k2chain}), the total number of vertices is $n=km$.  We 
obtain a generating function of the form given by eq.  (\ref{gammagen}) with
(\ref{n}) and (\ref{d}), where $deg_x({\cal N})=3$ and $deg_x({\cal D})=5$.
The denominator of the generating function is 
\beqs 
{\cal D}&=&[1-(-1)^kx](1-D_{k+2}x)[1+(-1)^k(q-2)x] \times \cr\cr & & \Biggl [1+
\biggl [D_{k+1}+(-1)^k(q-2) \biggr ] x + \biggl [D_{k+1} (-1)^k(q-2)
-(-1)^kD_k \biggr ] x^2 \Biggr ]
\eeqs 
We thus have 
\beq 
\lambda_{h,1}=(-1)^k 
\eeq
\beq 
\lambda_{h,2}=D_{k+2} 
\eeq 
\beq 
\lambda_{h,3}=-(-1)^k(q-2)
\eeq
\beq
\lambda_{h,(4,5)}=\frac{1}{2}\Biggl [ -D_{k+1}-(-1)^k(q-2)\pm (
[D_{k+1}-(-1)^k(q-2)]^2+4D_k(-1)^k)^{1/2} \ \Biggr ] 
\eeq 
(where the subscript $h$ means homeomorphic.) 
For the chromatic polynomial we obtain 
\beq
P((Ch)_{k,2,\Sigma_{+,m},cyc.,m},q) = (q^2-3q+1)(-1)^{km} + (\lambda_{h,2})^m +
(q-1)\Bigl [ (\lambda_{h,4})^m+(\lambda_{h,5})^m \Bigr ]
\label{phe}
\eeq
Note that $\lambda_{h,3}$ does not occur in $P$. 
For illustration, the coefficient functions entering in the numerator 
${\cal N}$ are given in the Appendix for $k=3,4$.  

Fig. \ref{cycgon56a} shows the analytic structure of the $W$ function for
a cyclic strip of the square lattice with homeomorphic expansion parameters
$(k_1,k_2)=$ (a) (2,3) (b) (2,4).  In these cases we find the respective values
(a) $q_c \simeq 2.455$ and (b) $q_c=2$. 
In the region $R_1$  the leading term is $\lambda_{h,2}$. In these cases the 
boundaries ${\cal B}$ do not have any part with $Re(q) < 0$. 
\begin{figure}
\centering
\leavevmode
\epsfxsize=3.5in
\begin{center}
\leavevmode
\epsffile{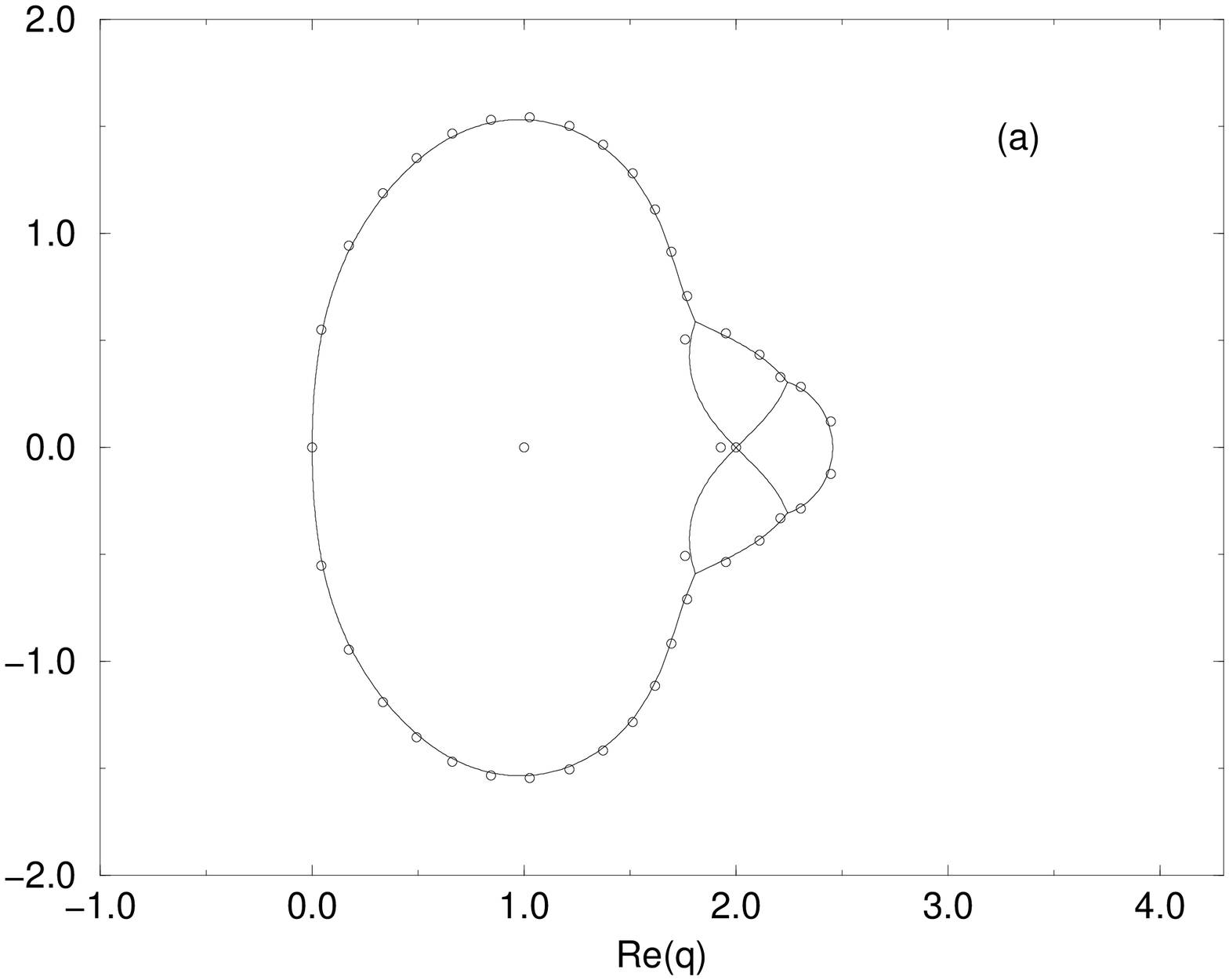}
\end{center}
\vspace{-4cm}
\begin{center}
\leavevmode
\epsfxsize=3.5in
\epsffile{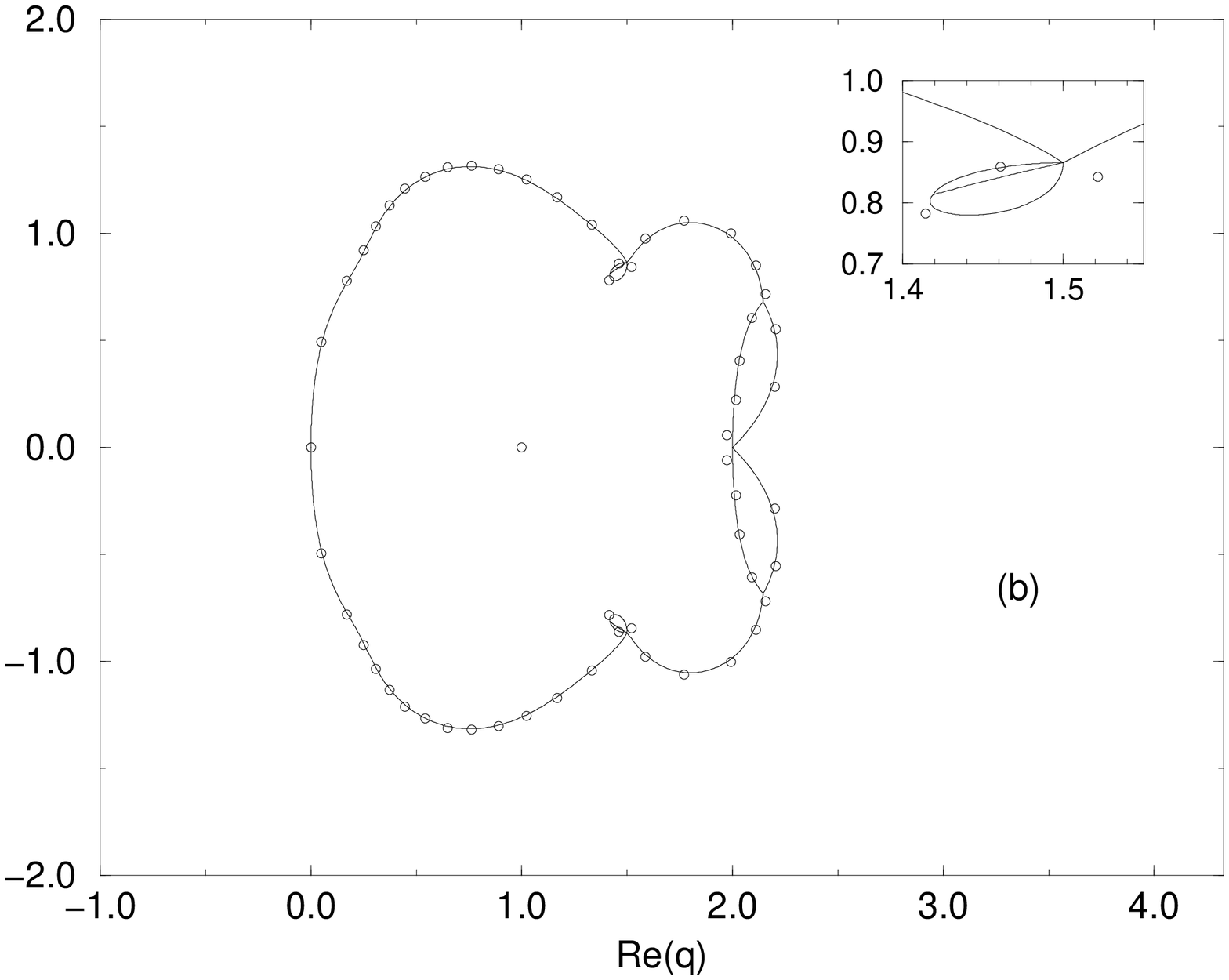}
\end{center}
\vspace{-2cm}
\caption{\footnotesize{Boundary ${\cal B}$ in the $q$ plane for 
$W$ function for an asymmetric cyclic chain of (a) $5$-gons ($k=3$) 
(b) $6$-gons ($k=4$). Chromatic zeros for $m=14$ are shown for comparison.}}
\label{cycgon56a}
\end{figure}

\section{Conclusions} 

In this paper, we have given exact calculations of the zero-temperature
partition function of the $q$-state Potts antiferromagnet strips of the square,
kagom\'e and triangular lattices with periodic longitudinal boundary conditions
(and free transverse boundary conditions).  These yield, in the limit of
infinite length, the ground-state entropy $S_0=k_B \ln W$ via eq.
(\ref{w}). These results are of fundamental interest since this model exhibits
nonzero ground state entropy $S_0 > 0$ for sufficiently large $q$ and hence is
an exception to the third law of thermodynamics.  We also include results for
homeomorphic expansions of the square lattice strip.  The analytic properties
of $W(q)$ are determined and related to zeros of the chromatic polynomial for
long finite strips.

\vspace{10mm}

This research was supported in part by the NSF grant PHY-97-22101.

\newpage

\section{ Appendix}

We list in this appendix the coefficient functions for the numerators of the
generating functions for the cyclic strips of the square and kagom\'e lattices
and for asymmetric homeomorphic expansions of the cyclic square strip. 

\subsection{Strip of the Square Lattice}

For the cyclic strip of the square lattice with $L_y=3$ we calculate
\beq
A_{sq,0} = q(q-1)(q^2-3q+3)^2
\eeq
\beq
A_{sq,1} = -2q(q-1)(q^5-6q^4+11q^3-21q+16)
\eeq
\beqs
A_{sq,2} &=& -q(q-1)(3q^8-45q^7+306q^6-1242q^5+3301q^4-5854q^3 \cr\cr
& &+6703q^2-4482q+1318)
\eeqs
\beqs
A_{sq,3} &=& -2q(q-1)(q^{10}-24q^9+250q^8-1516q^7+5992q^6-16230q^5 \cr\cr
& &+30580q^4-39552q^3+33530q^2-16764q+3732)
\eeqs
\beqs
A_{sq,4} &=& q(q-1)(7q^{11}-159q^{10}+1636q^9-10077q^8+41328q^7-118548q^6\cr\cr
& &+242569q^5-353514q^4+358776q^3-240824q^2+95932q-17119)
\eeqs
\beqs
A_{sq,5} 
&=& -2q(q-1)^2(4q^{11}-91q^{10}+935q^9-5728q^8+23249q^7-65620q^6\cr\cr
& &+131260q^5-185622q^4+181164q^3-115604q^2+43126q-7070)
\eeqs
\beqs
A_{sq,6}
&=& q(q-1)^2(2q^{12}-49q^{11}+534q^{10}-3412q^9+14151q^8-39685q^7\cr\cr
& &+75440q^6-92942q^5+62238q^4-455q^3-36019q^2+27008q-6808)
\eeqs
\beqs
A_{sq,7} &=& 2q(q-2)(q-3)(q-1)^4(q^9-19q^8+162q^7-818q^6+2706q^5\cr\cr
& &-6090q^4+9318q^3-9330q^2+5535q-1477)
\eeqs
\beqs
A_{sq,8} &=& -q(q-4)(q-2)^3(q-1)^5(q^3-6q^2+13q-11) \times \cr\cr
& &(q^4-9q^3+29q^2-40q+22) 
\eeqs

\subsection{Strip of the Kagom\'e Lattice}

For our cyclic strip of the Kagom\'e (kg) lattice we find 
\beq
A_{kg,0} = q(q-1)^2(q-2)^2
\eeq
\beq
A_{kg,1} = -q(q-1)(q-2)^2(2q^5-17q^4+59q^3-105q^2+104q-59)
\eeq
\beqs
A_{kg,2} &=& q(q-1)(q-2)(2q^9-37q^8+311q^7-1547q^6+4994q^5 \cr\cr
& &-10845q^4+15932q^3-15484q^2+9250q-2656)
\eeqs
\beqs
A_{kg,3} &=& q(q-1)(q-2)(3q^{11}-69q^{10}+726q^9-4604q^8+19522q^7-58029q^6 
\cr\cr 
& &+123231q^5-186853q^4+198437q^3-141104q^2+60968q-12292)
\eeqs
\beqs
A_{kg,4}
&=&-q(q-1)^2(q-2)^2(2q^{11}-34q^{10}+251q^9-1020q^8+2290q^7-1840q^6-4218q^5
\cr\cr 
& &+14163q^4-16324q^3+4750q^2+6204q-4476)
\eeqs
\beqs
A_{kg,5} &=& q(q-1)^2(q-2)^4(7q^{10}-121q^9+953q^8-4521q^7+14396q^6-32498q^5
\cr\cr
& &+53581q^4-65213q^3+57416q^2-33432q+9720)
\eeqs
\beqs
A_{kg,6} &=& -q(q-1)^3(q-2)^6(9q^8-129q^7+810q^6-2927q^5+6708q^4-10133q^3
\cr\cr
& & +10142q^2-6446q+2110)
\eeqs
\beq
A_{kg,7} = q(q-1)^4(q-2)^8(5q^6-54q^5+232q^4-511q^3+608q^2-380q+118)
\eeq
\beq
A_{kg,8} = -q^2(q-1)^6(q-2)^{11}(q-4)
\eeq

\subsection{Asymmetric Homeomorphic Expansions of the Square Strip}

For the case $k=3$ defined in the text, the coefficients entering in the 
numerator of the generating function are
\beq
A_{k=3,0} = q(q-1)(q-2)(q^3-4q^2+7q-5)
\eeq
\beq
A_{k=3,1} =-q(q-1)(q-2)(2q^4-12q^3+27q^2-29q+13)
\eeq
\beq
A_{k=3,2} =-q(q-1)^2(q-2)^2(q^2-q-1)
\eeq
\beq
A_{k=3,3} =q(q-2)^4(q-1)^2(q^2-2q+2)
\eeq

For the case $k=4$ we find 
\beq
A_{k=4,0} =q(q-1)(q^6-8q^5+28q^4-56q^3+70q^2-53q+19)
\eeq
\beq
A_{k=4,1} =q(q-1)(2q^7-21q^6+94q^5-236q^4+365q^3-356q^2+207q-56)
\eeq
\beq
A_{k=4,2} =-q(q-1)^2(2q^6-17q^5+59q^4-107q^3+107q^2-56q+11)
\eeq
\beq
A_{k=4,3} =-q(q-2)(q-1)^2(q^3-5q^2+8q-5)(q^4-5q^3+10q^2-10q+5)
\eeq

\vfill
\eject

\end{document}